\documentclass[a4paper]{jpconf}
\usepackage{graphicx}
\begin{document}
\title{Challenges for $\Lambda$CDM and MOND}

\author{Benoit Famaey}
\address{Observatoire Astronomique de Strasbourg, CNRS UMR 7550, France}

\author{Stacy McGaugh}
\address{Case Western Reserve University, USA}

\ead{benoit.famaey@astro.unistra.fr, ssm69@case.edu}

\begin{abstract}
The Universe on large scales is well described by the $\Lambda$CDM cosmological model. There however remain some heavy clouds on our global understanding, especially on galaxy scales, which we review here. While some of these clouds might perhaps disappear through small compensatory adjustments of the model, such as changing the mass of the dark matter particles or accounting better for baryonic physics, others should rather be taken as strong indications that the physics of the dark sector is, at the very least, much richer and complex than currently assumed, and that our understanding of gravity and dynamics might also be at play. For instance, the empirically well-tested MOND phenomenology in galaxies, whatever its final explanation, should be understood in any model of galaxy formation and dynamics. Current alternatives to $\Lambda$CDM however bring with them many unsolved questions and challenges.
\end{abstract}

\section{Introduction}
Assuming General Relativity to be valid on all scales, data ranging from the Cosmic Microwave Background to individual galaxies point towards a Universe dominated by dark energy and dark matter, the nature of these being most certainly amongst the deepest problems of modern physics. While dark energy is well-represented by a cosmological constant $\Lambda$ in Einstein's field equations, the currently preferred dark matter candidate is a collection of stable, neutral, elementary particles that condensed from the thermal bath of the early Universe, and which are known as `cold dark matter' (CDM) particles (see, e.g., Frenk \& White 2012, Strigari 2012 for recent reviews). On galaxy scales, however, predictions of this standard $\Lambda$CDM cosmological model, although plagued by the enormous complications of baryonic astrophysics, are difficult to reconcile with observations. We hereafter review these challenges for the $\Lambda$CDM model, and point out that some of them hint at a richer and more complex physics of the dark sector than currently assumed. In particular, many observed scaling relations involve the ubiquitous appearance of an acceleration constant $a_0 \approx \Lambda^{1/2} \sim 10^{-10} {\rm m/s}^2$, whose origin is a deep mystery in the standard context. Surprisingly, most of these scaling relations can be summarized by the empirical formula of Milgrom (1983). The success of this formula means that the gravitational field in galaxies mimicks, for whatever reason, an effectively modified force law on galaxy scales, known as Modified Newtonian Dynamics (MOND). This formula however fails to account for dynamics and lensing of galaxy clusters (e.g., Clowe et al. 2006, Angus et al. 2007), meaning that if the formula would be reflecting a true modification of gravity as an alternative to galactic dark matter, it should rely at cluster scales on residual missing mass, which could be in baryonic or non-baryonic form (Milgrom 2008, Angus et al. 2007), or the formula should be extended (Zhao \& Famaey 2012). It is also unclear how the angular power spectrum of the Cosmic Microwave Background (CMB) could be accounted for without resorting to some form of non-baryonic dark matter (e.g., Slosar et al. 2005, Angus 2009). Nevertheless, the main motivation for studying alternatives to $\Lambda$CDM based on the MOND phenomenology is \textit{not} necessarily to get rid of any form of `dark matter', but rather to explain why the observed gravitational field in galaxies is apparently mimicking a universal force law generated by the baryons alone. The simplest explanation is of course {\it a priori} not that dark matter particles arrange themselves (by a hitherto unknown physical mechanism) in order to mimick a fake force law, but rather that the force law itself is modified. However, at a fundamental level, covariant theories of modified gravity often have to include new fields in the dark sector to reproduce this effective force law (fields with an energy density nevertheless subdominant to the baryonic one, and a role completely different from that of CDM: they would mediate the force rather than sourcing it), or even introduce what could be called a `dark matter medium' (with an energy density outweighing the baryonic one) exhibiting a kind of new fundamental interaction with baryons: this makes the confrontation between MOND and dark matter much less clear than often believed, since the former implies a more complex structure of the dark sector than the currently assumed CDM, but does not necessarily imply the absence of a dark sector. In MOND, the new fields responsible for the effects usually attributed to CDM would in fact be somewhat closer to dark energy fields than to CDM. It would of course be even more exciting if one would manage to find a physical connection between these putative new fields and the actual dark energy sector. 

We herefater list a (non-exhaustive) series of problems for $\Lambda$CDM (Sect.~2), then show how theories based on the MOND phenomenology might address a lot of them (Sect.~3), before listing a (non-exhaustive) series of covariant theories currently proposed in this vein (Sect.~4), together with their own questionmarks and internal problems. The reading of this short proceeding can be completed by the reading of the exhaustive review paper recently published in Living Reviews in Relativity (Famaey \& McGaugh 2012).

\section{Current problems for our understanding of galaxy properties in $\Lambda$CDM}

\vspace*{2mm}

\subsection{{\bf The nature and distribution of dwarf galaxies}}

\begin{itemize}
\item[{\it 2.1.a.}] \textbf{The missing satellites challenge.} $\Lambda$CDM simulations predict vast numbers of subhaloes that are satellites to the main halo hosting a galaxy like the Milky Way.  One would naively expect each of these subhaloes to host their own miniature satellite galaxy.  The Local Group looks nothing like this prediction, having only a small handful of dwarfs around each of the giant galaxies. Taking into account stellar feedback and heating processes (mainly at re-ionisation) in the galaxy formation process, the predicted number of faint satellites around a Milky Way-like galaxy is $\sim$~100 to 600. Since the majority of the 24 known satellites of the Milky Way have been largely discovered with the Sloan Digital Sky Survey (SDSS), and since this survey covered only one fifth of the sky, it has been argued that the problem was solved. However, models that successfully explain the properties of the Milky Way dwarf satellites predict unobserved dwarfs beyond the virial radius (Bovill \& Ricotti 2011a).  Moreover, the models tend to produce an overabundance of bright dwarf satellites ($L_V > 10^4 L_\odot$) with respect to observations (Bovill \& Ricotti 2011b). A rather discomforting way-out is to simply state that the Milky Way must be a statistical outlier, but this is actually contradicted by the study of Strigari \& Wechsler (2012) on the abundance of bright satellites around Milky Way-like galaxies in the SDSS survey. A somewhat related but nevertheless distinct problem is that simulations predict that the most massive subhaloes of the Milky Way are too {\it dense} to host any of the observed bright satellite galaxies (Boylan-Kolchin et al. 2011, 2012). This is the  `too big to fail' aspect of the missing satellites challenge. A potential solution might be warm rather than cold dark matter.

\item[{\it 2.1.b.}] \textbf{The satellites phase-space correlation challenge (disks of satellites).} The distribution of dark subhalos around Milky Way-sized halos is also predicted by $\Lambda$CDM to be roughly isotropic. However, the Milky Way satellites are currently observed to be highly correlated in phase-space: they lie within a seemingly rotationally supported, relatively thin disk (Fig.~1, see Kroupa et al. 2010).  This differs from the prediction for subhalos in both coordinate and momentum space. Since the SDSS survey covered only one fifth of the sky, it will of course be most interesting to see whether ongoing surveys such as Pan-STARRS will confirm this state of affairs. Whether or not such a satellite phase-space correlation would be unique to the Milky Way should also be carefully checked: the Milky Way could be a statistical outlier, but if the $\Lambda$CDM model is a realistic description of nature, then the average satellite configurations in external galaxies should be only moderately flattened, and generally non-rotating. In this respect, the recent findings of Ibata et al. (2013) that half of the satellites of M31 define an extremely thin and extended rotating structure makes this phase-space correlation challenge even more severe (see Sect.~3 for a possible solution). 
\begin{figure}[h]
\includegraphics[width=18pc]{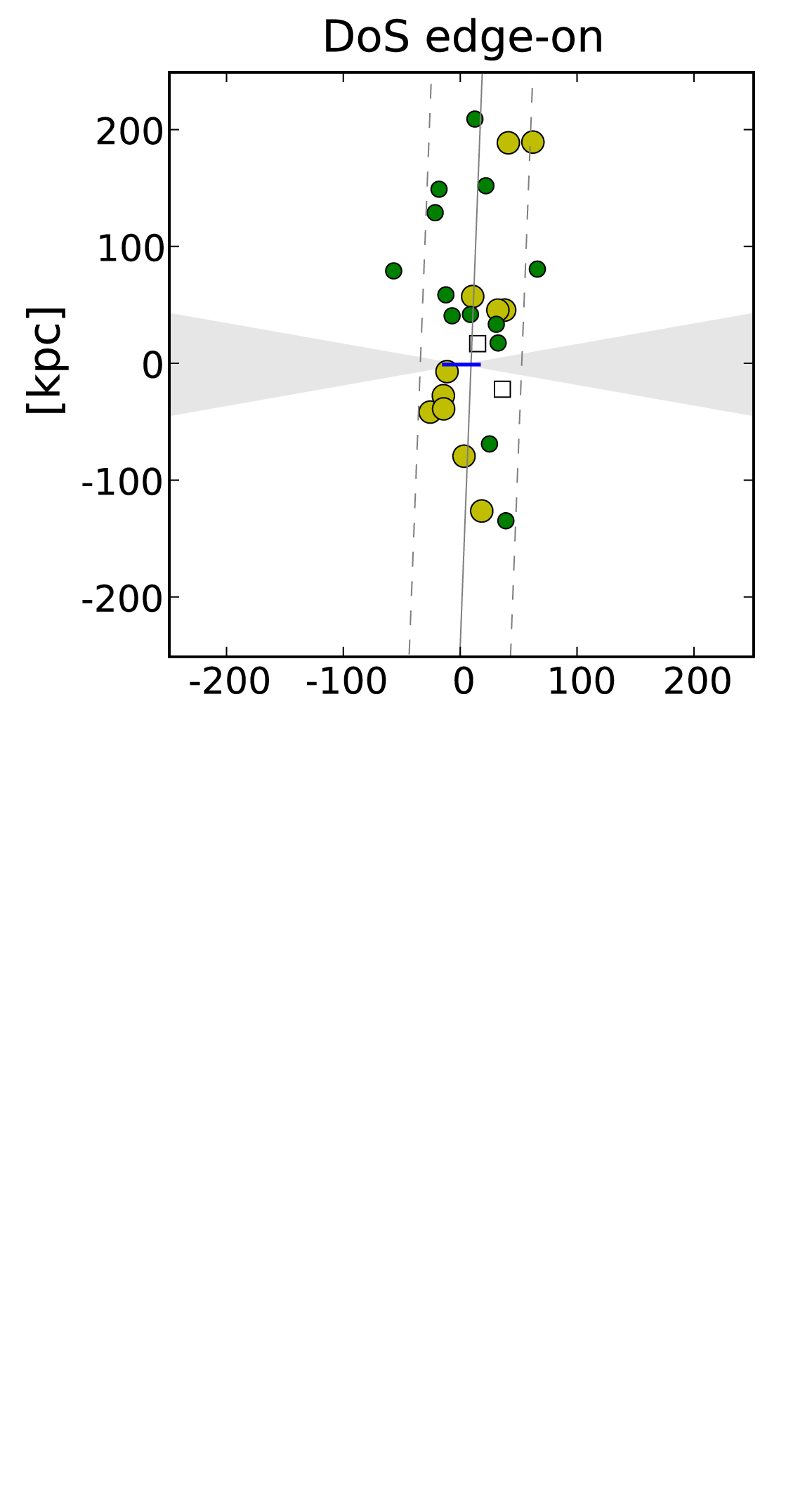}\hspace{2pc}%
\begin{minipage}[b]{16pc}\caption{\label{label1}Spatial distribution of the Milky Way (MW) satellites (from Kroupa et al. 2010). Both axes in kpc. The 11 classical satellites are shown as large (yellow) circles, the 13 recently discovered satellites are represented by the smaller (green) dots, Pisces I and II being the two southern dots. The two open squares near the MW are Seg 1 and 2. The obscuration-region from the MW disk is given by the horizontal gray area.}
\end{minipage}
\end{figure}

\item[{\it 2.1.c.}] \textbf{The density-morphology relation of dwarf ellipticals.} More dwarf elliptical galaxies are observed in denser environments (e.g. Kroupa et al. 2010). This relation, observed in the field, in galaxy groups and in galaxy clusters, is not yet understood.
\end{itemize}

\subsection{{\bf The dynamical friction and stabilization from dark matter}}

\begin{itemize}
\item[{\it 2.2.a.}] \textbf{The angular momentum challenge.}  Both the merger history of galaxy disks in a hierarchical formation scenario and the associated transfer of angular momentum from the baryonic disk to the dark halo cause the specific angular momentum of the baryons to end up being much too small in simulated disks. These simulated disks in turn end up much smaller than the observed ones, and elliptical systems end up too concentrated too. To address this challenge within the standard paradigm, one needs to form disks through late-time quiescent gas accretion from large-scale filaments, with less late-time mergers than currently predicted in $\Lambda$CDM.

\item[{\it 2.2.b.}] \textbf{The pure disk challenge.} Related to the previous problem, large \textit{bulgeless} thin disk galaxies are very difficult to produce in simulations. This is because major mergers typically create bulges. This is fundamental to the hierarchical nature of galaxy formation in $\Lambda$CDM, wherein galaxies are built by the merger of smaller galaxies.  Bulgeless galaxies should therefore represent the rare, quiescent tail of a distribution of merger histories for galaxies of the Local Volume. However, these bulgeless disk galaxies are quite common, representing more than half of large galaxies (with $V_c > 150\mathrm{\ km/s}$) in the Local Volume. 

\item[{\it 2.2.c.}] \textbf{The stability challenge.}  Quasi-spherical CDM halos stabilize very low surface density disks against the formation of bars and spirals, due to a lack of disk self-gravity. The observation (McGaugh et al. 1995) of Low Surface Brightness (LSB) disk galaxies with prominent bars and spirals is thus challenging in the absence of a significant disk component of dark matter. Moreover, in the absence of such a disky dark matter component, the lack of disk self-gravity prevents the creation of observed large razor thin LSB disks. This suggests more disk self-gravity than meets the eye (McGaugh \& de Blok 1998). In the standard context, this would tend to point towards an additional disk-specific DM component, either a CDM-one linked to in-plane accretion of satellites, or a baryonic one, for instance in the form of molecular gas or due to a bottom-heavy IMF.
\end{itemize}

\subsection{{\bf Structure formation not fast enough in $\Lambda$CDM?}}

\begin{itemize}
\item[{\it 2.3.a.}] \textbf{The bulk flow challenge.} In the $\Lambda$CDM model, peculiar velocities of galaxy clusters are predicted to be of order $\sim 200$~km/s. These velocities can actually be measured through the fluctuations in the CMB generated by the scattering of the CMB photons by the hot X-ray-emitting gas inside galaxy clusters. These observations yield an observed coherent bulk flow of order $\sim 1000$~km/s  on scales out to at least 400~Mpc (Kashlinsky et al. 2012). This bulk flow challenge also appears in galaxy studies (Watkins et al. 2009). A related problem is the collision velocity of the merging bullet cluster 1E0657-56 at $z=0.3$, which is larger than 3100~km/s, too high a value to be accounted for by $\Lambda$CDM. These observations seem to indicate that the attractive force is enhanced compared to what $\Lambda$CDM predicts.

\item[{\it 2.3.b.}] \textbf{The high-z clusters challenge.} The observation of a single very massive cluster at high redshift can falsify $\Lambda$CDM. In this respect the existence of galaxy clusters like El Gordo with a mass of $\sim 2 \times 10^{15}\,M_{\odot}$ at $z = 0.87$ (Menanteau et al. 2012) and XMMU~J2235.3-2557 with a mass of of $\sim 4 \times 10^{14}\,M_{\odot}$ at $z = 1.4$ is surprising.  Though not sufficient to rule out the model, these clusters certainly push the envelope: there is only about a 50:50 chance that a cluster as massive as El Gordo exists at such high redshift.  This has been the empirical experience: for every generation of surveys, structures are found further away, and appear sooner than expected.

\item[{\it 2.3.c.}] \textbf{The Local Void challenge.} The 562 known galaxies at distances smaller than 8~Mpc from the center of the Local Group define the `Local Volume'. Within this volume, the region known as the `Local Void' hosts only 3 galaxies. This is much less than the expected $\sim 20$ galaxies for a typical similar void in the $\Lambda$CDM model (Peebles \& Nusser 2010). Moreover, in the Local Volume, large luminous galaxies are 6 times more abundant than predicted in the underdense regions. This could just mean that the Local Volume is a statistical outlier, but it could also point, in line with the two previous problems, towards more rapid structure formation in the early Universe, allowing sparse regions to form large galaxies cleaning their environment in a shorter time, thus making the voids emptier and the galaxies larger at early times.
\end{itemize}

\subsection{{\bf Systematics in the dark-to-baryonic matter ratios and the ubiquitous $a_0$-scale}}

\begin{itemize}

\item[{\it 2.4.a.}] \textbf{The cusp-core challenge.}  A long-standing problem of $\Lambda$CDM is the fact that the numerical simulations of the collapse of dark matter halos lead to a `cuspy' density distribution that rises steeply as a decreasing function of radius, while rotation curves of external galaxies imply nearly constant density cores in the central parts. The state-of-the-art solution to this problem is to enforce strong supernovae outflows that move large amounts of low-angular-momentum gas from the central parts and that pull on the central dark matter concentration to create a core, but this is still a relatively fine-tuned process, which fails to account for cored profiles in the smallest galaxies, and fails to produce their observed baryon fractions

\item[{\it 2.4.b.}] \textbf{The missing baryons challenge(s): baryon fraction defined by the acceleration in units of $a_0$.}  Constraints from the CMB imply $\Omega_m = 0.27$ and $\Omega_b=0.046$. However, our inventory of known baryons in the local Universe comes up short of the total. For example, Bell et al. (2003) estimate that the sum of cold gas and stars is only $\sim$~5\% of $\Omega_b$. While it now seems that many of the missing baryons are in the form of highly ionized gas in the warm-hot intergalactic medium (WHIM), we are still unable to give a confident account of where all the baryons reside. But there is another missing baryons challenge within each CDM halo: one would indeed naively expect each halo to have the same baryon fraction as the Universe, $f_b=\Omega_b/\Omega_m=0.17$. On the scale of clusters of galaxies, this is approximately true (but still systematically low), but for galaxies, observations depart from this in a systematic way which remains unexplained in the standard context. In Fig.~2 (McGaugh et al. 2010), the ratio of the detected baryon fraction over the cosmological one, $f_d$, is plotted as a function of the potential well of the systems. There is a clear correlation: at large radii, the baryon fraction is actually directly equivalent to the acceleration in units of $a_0 = 10^{-10} {\rm m/s}^2$. If we adopt a rough relation $M_{500} \simeq 1.5 \times 10^{5}\ M_{\odot} \times V_c^3 ({\rm km/s})^{-3}$, we get that the acceleration at $R_{500}$, and thus the baryon fraction, is $M_b/M_{500} = a_{500}/a_0 \simeq 4 \times 10^{-4} \times V_c ({\rm km/s})^{-1}$. Divided by the cosmological baryon fraction, this explains the trend for $f_d= M_b/(0.17 M_{500})$ with potential ($\Phi = V_c^2$) in Fig.~2. This missing baryons challenge is actually closely related to the baryonic Tully--Fisher relation.

\begin{figure}[h]
\includegraphics[width=16pc]{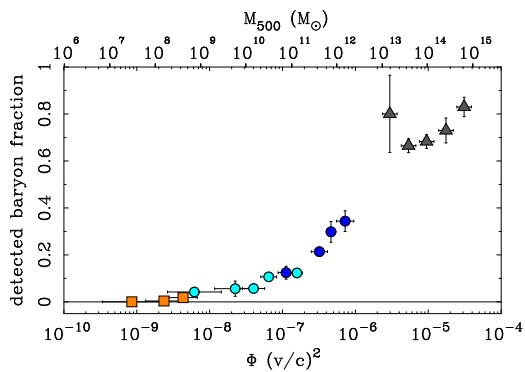}\hspace{2pc}%
\begin{minipage}[b]{14pc}\caption{\label{label2}The fraction of detected over expected baryons $f_d = (M_b/M_{500})/f_b$ as a function of potential well and mass. Each point represents many objects. Orange squares are Local Group dwarf satellites, light blue circles gas-dominated disk galaxies, dark blue circles star-dominated spiral galaxies, grey triangles galaxy clusters.}
\end{minipage}
\end{figure}

\item[{\it 2.4.c}] \textbf{The baryonic Tully--Fisher relation (with $Ga_0$ defining zero point).} The baryonic Tully--Fisher relation is an observed relation between the asymptotic circular velocity and baryonic mass in galaxies (Fig.~3). Recent work extending the relation to low mass, typically low surface brightness and gas rich galaxies, extends the dynamic range of the relation to five decades in baryonic mass. Over this range, the relation has remarkably little intrinsic scatter (consistent with zero given the observational errors) and is well described as a power law, or equivalently, as a straight line in log-log space ${\rm log} M_b = \alpha {\rm log} V_f - {\rm log} \beta$, with slope $\alpha = 4$. This slope is consistent with a constant acceleration scale $a_0 \approx V_f^4/(GM_b)$ such that the normalization constant $\beta = Ga_0$. All rotationally-supported galaxies are observed to follow the relation, irrespective of their size, surface brightness, or gas fraction.

\item[{\it 2.4.d}] \textbf{Tidal dwarf galaxies.} Even young tidal dwarf galaxies (TDGs) formed in the collision of larger galaxies appear to obey the baryonic Tully--Fisher relation (Gentile et al. 2007). This is surprising because these galactic collisions should be very effective at segregating dark and baryonic matter: the rotating gas disks of galaxies that provide the fodder for tidal tails and the TDGs that form within them initially have nearly circular, coplanar orbits. In contrast, the dark matter particles are on predominantly radial orbits in a quasi-spherical distribution. This difference in phase space leads to tidal tails that themselves contain very little dark matter. When TDGs form from tidal debris, they should be largely devoid of dark matter. Nevertheless, TDGs do appear to contain enough dark matter to obey the same baryonic Tully-Fisher relation as galaxies that have not suffered this baryon-dark matter segregation process.

\begin{figure}[h]
\includegraphics[width=14pc]{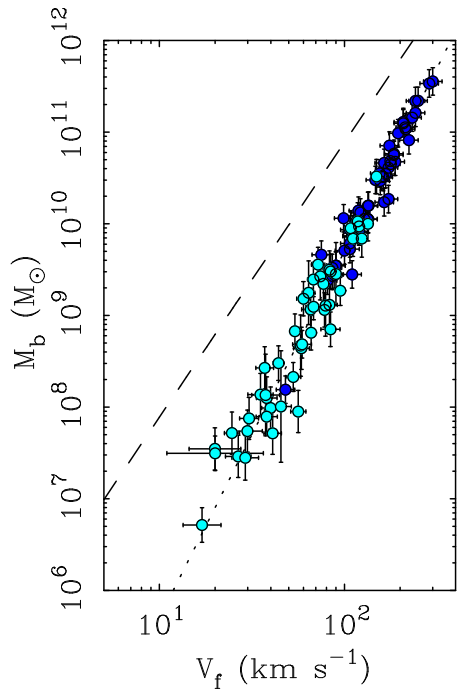}\hspace{2pc}%
\begin{minipage}[b]{14pc}\caption{\label{label3}The Baryonic Tully--Fisher
    relation for galaxies with well-measured asymptotic velocities $V_f$. The dark blue points have $M_* > M_g$, while the light blue points have $M_* < M_g$ and are generally less precise in velocity, but more accurate in terms of mass. The dotted line has the observed slope 4, the dashed line has slope 3, the difference between these two lines being the origin of the variation in the detected baryon fraction in Fig.~2.}
\end{minipage}
\end{figure}

\item[{\it 2.4.e.}] \textbf{$a_0$ as a transition acceleration.} The mass discrepancy in galaxies always appears (transition from baryon dominance to dark matter dominance) when $V_c^2/R \sim a_0$, yielding a clear mass-discrepancy acceleration relation (Fig.~4). This, again, is the case for every single rotationally supported system irrespective of its formation mechanism and history. For High Surface Brightness (HSB) galaxies, where there exists two distinct regions where $V_c^2/R>a_0$ in the inner parts and $V_c^2/R<a_0$ in the outer parts, locally measured mass-to-light ratios show no indication of hidden mass in the inner parts, but rise beyond the radius where $V_c^2/R \approx a_0$. It can never be emphasized enough that the role played by $a_0$ in the zero-point of the baryonic Tully-Fisher relation and this role as a transition acceleration have strictly no intrinsic link with each other, they are fully independent of each other within the CDM paradigm. There is nothing built in $\Lambda$CDM that stipulates that these two relations (the existence of a transition acceleration and the baryonic Tully-Fisher relation) should exist at all, and of course nothing that implies that these should exhibit an identical acceleration scale.

\begin{figure}[h]
\includegraphics[width=20pc]{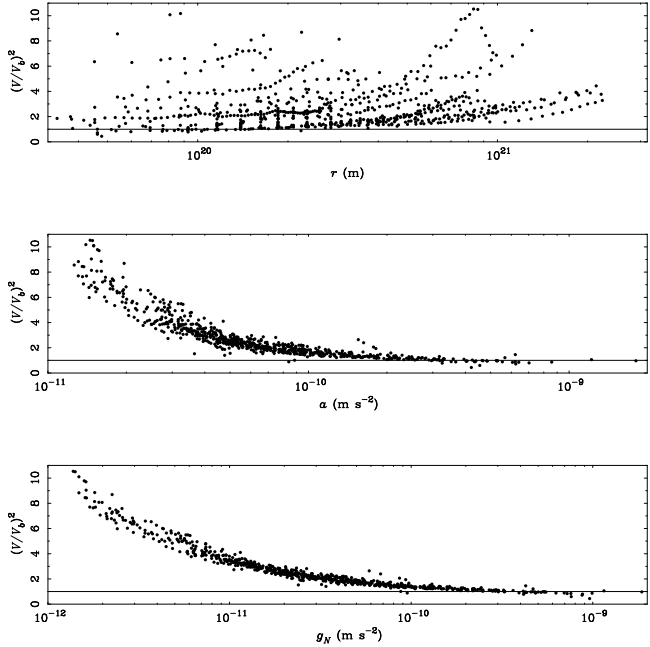}\hspace{2pc}%
\begin{minipage}[b]{14pc}\caption{\label{label4}The mass discrepancy
  is defined as $(V/V_b)^2$ where $V$ is the observed velocity and
  $V_b$ is the velocity attributable to baryonic matter.  
  Many hundreds of individual resolved measurements along the rotation curves  
  of $\sim 100$ spiral galaxies are plotted.
  The mass discrepancy is plotted as a function
  of radius (top panel), centripetal acceleration $a = V^2/r$ (middle panel), and Newtonian acceleration 
  $g_N = V_b^2/r$ (bottom panel).}
\end{minipage}
\end{figure}

\begin{figure}[h]
\includegraphics[width=18pc]{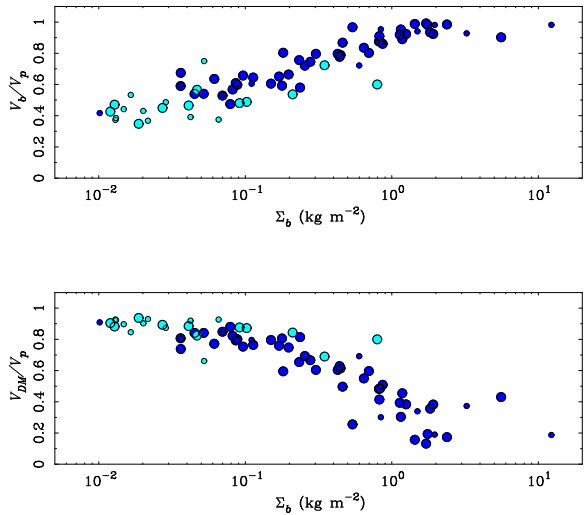}\hspace{2pc}%
\begin{minipage}[b]{14pc}\caption{\label{label5}The fractional contribution of baryons ($V_b/V_p$, top) and dark matter ($V_{DM}/V_p$, bottom) to the rotation velocity $V_p$ at the radius $R_P$ where the contribution of the baryons peaks, plotted as a function of the characteristic surface density ($\Sigma_b = 0.75 M_ b /R_p^2$). As the baryonic surface density increases, the fractional contribution of dark matter to the total gravitating mass decreases. This fine-tuning persists for any choice of stellar mass-to-light ratio.}
\end{minipage}
\end{figure}

\item[{\it 2.4.f.}] \textbf{$a_0/G$ as a transition central surface density.} The acceleration $a_0$ defines the transition from LSB galaxies to HSB galaxies: in LSB galaxies whose {\it central} surface density is much smaller than some critical value of order $\Sigma_{\dagger}=a_0/G$, DM dominates everywhere, and the magnitude of the mass discrepancy is given by the inverse of the acceleration in units of $a_0$. On the other hand, baryons dominate in the inner parts of HSB galaxies whose central surface density is higher than $\Sigma_{\dagger}$. There is an anti-correlation between the baryonic surface density and the fractional contribution of DM to the rotation curve (Fig.~5). To explain this, there must be a strong fine-tuning between dark and baryonic surface densities, a sort of repulsion between them, a repulsion which is however contradicted by the correlations between baryonic and dark matter bumps and wiggles in rotation curves (known as `Renzo's rule'). The shapes of rotation curves also depend on surface density: HSB galaxies have rotation curves that rise steeply then become flat, or even fall somewhat to the not-yet-reached asymptotic flat velocity, while LSB galaxies have rotation curves that rise slowly to the asymptotic flat velocity. Finally the \textit{total} (baryons+DM) acceleration declines with the mean \textit{baryonic} surface density of galaxies, in the form $a \propto \Sigma_b^{1/2}$ (Fig.~6).

\item[{\it 2.4.g.}] \textbf{Features in the baryonic distribution imply features in the rotation curve.} This is known as `Renzo's rule' (Sancisi 2004, see also Swaters et al. 2012). While the effect of non-axisymmetric motions should be investigated in more detail, this is currently most easily interpreted by either a heavy `dark baryonic' component (e.g., in the form of molecular H2) scaling with HI in galaxy disks, or by a theory where the baryons effectively act as the main source of gravity.

\item[{\it 2.4.h.}] \textbf{$a_0/G$ as a critical mean surface density for stability.} Disks with mean baryonic surface density $\langle \Sigma \rangle > \Sigma_\dagger=a_0 /G$ are extremely rare (Fig.~7) and unstable.  If stability is provided by a DM halo, there is no reason for this stability threshold to be related to the same acceleration scale that appears elsewhere (so this is again an independent occurence of the $a_0$-scale). Dense halos predicted by $\Lambda$CDM could (and should) in principle host higher surface brightness disks than are observed.

\begin{figure}[h]
\includegraphics[width=18pc]{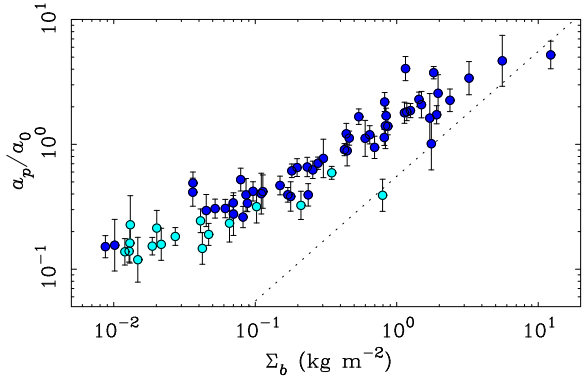}\hspace{2pc}%
\begin{minipage}[b]{14pc}\caption{\label{label6}The characteristic centripetal acceleration $a_p = V_p^2/R_p$ in units of $a_0$ plotted against the characteristic baryonic surface density. The data do show a correlation ($a_p \propto \Sigma_b^{1/2}$), clearly indicating a dynamical role for the baryons, whilst CDM dominance should normally erase any such correlation.}
\end{minipage}
\end{figure}

\begin{figure}[h]
\includegraphics[width=20pc]{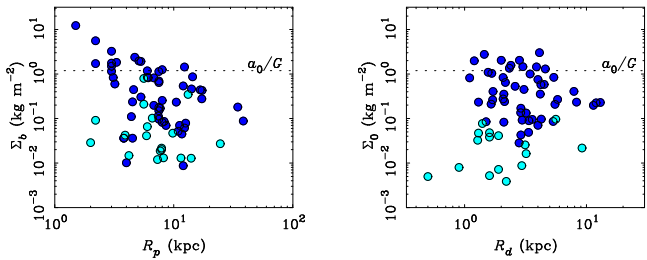}\hspace{2pc}%
\begin{minipage}[b]{14pc}\caption{\label{label7}The characteristic surface density of baryons vs. their dynamical scale length $R_p$ (Left), and disk-only central surface density vs. disk scale-length (Right). High characteristic surface densities at low $R_p$ in the left panel are typical of bulge-dominated galaxies, and disks with characteristic surface density $>a_0/G$ are extremely rare.}
\end{minipage}
\end{figure}
\end{itemize}

\section{Solving problems: Milgrom's law and MOND}

As we have seen, the $a_0$ acceleration-scale is ubiquitous in galaxies. It defines the zero point of the Tully-Fisher relation and the amplitude of the mass-discrepancy in the weak-gravity limit,  it appears as a transition-acceleration above which no DM is needed, and below which DM appears, and it defines the maximum surface density of pure disks.  It would obviously be interesting if these observational occurences could be summarized and empirically unified in some way: such a unification \textit{is} actually feasible through the rather old idea of Milgrom (1983) that the total gravity ${\bf g}$ is related to the baryon-generated one ${\bf g_N}$ in a way analogous to Coulomb's law in a dieletric medium:
\begin{equation}
\mu\left(\frac{g}{a_0}\right){\bf g}={\bf g_N}, 
\label{moti}
\end{equation}
where the interpolating function
\begin{equation}
\mu(x) \rightarrow 1 \; {\rm for} \; x \gg 1 \; {\rm and} \; \mu(x) \rightarrow x \; {\rm for} \; x \ll 1.
\label{muasympt}
\end{equation}
This law is at the basis of the Modified Newtonian Dynamics (MOND) paradigm. It does \textit{not} simply make rotation curves flat, but predicted {\it all} the challenging observations of Sect.~2.4 in a quick and efficient way (see Sect.~5.2 of Famaey \& McGaugh 2012). This formula was not constructed to explain already known observational facts, but predicted almost all of them {\it a priori}. Its salient properties can be elegantly summarized by simply stating that the underlying theory, whatever its deep nature, has to become scale-invariant under transformations $(t,{\bf r}) \rightarrow \lambda(t, {\bf r})$ in the regime of small acceleration $a \ll a_0$ (see, e.g., Milgrom 2012).

Addressing all the other challenges, from Sect.~2.1 to 2.3, would need a full-fledged theory predicting Milgrom's law and the associated MOND phenomenology. As we shall see in the next section, such a full-fledged theory does not exist at the moment. One can however speculate on how these problems could be addressed by such a theory. For instance, if the fields generating the MOND behavior on galaxy scales have an energy density subdominant to the baryonic one, effects such as dynamical friction with DM would disappear. Galaxy disks would then easily get more extended thanks to the absence of angular momentum transfer to the dark halo, but mergers would also take longer, and be subject to multiple passages (Tiret \& Combes 2008). While MOND would naturally provide added stability compared to Newtonian dynamics without DM, it does not over-stabilize disks: features like bars and spiral arms are a natural result of disk self-gravity,  and numerical simulations based on MOND appear to do a good job at reproducing the range of observed morphologies of spiral galaxies (Tiret \& Combes 2007, 2008). Concerning structure formation, the MOND paradigm has {\it a priori} no cosmology, providing analogs for neither the Friedmann equation nor the Robertson-Walker metric. For these, one must appeal to specific hypotheses for the covariant parent theory of MOND (Sect.~4), which is far from unique. What is more, none of the candidates is theoretically satisfactory, as none emerges from first principles. It is thus not clear whether a compelling candidate MOND cosmology will ever emerge. One can however speculate on what would be a reasonable scenario, based on simulations using MOND in the weak-field regime at their basis (e.g., Sanders 1998, Angus \& Diaferio 2011). In the absence of CDM, and with a MOND-like force driving structure formation, after initially lagging behind, structure formation would rapidly speed up compared to $\Lambda$CDM once the influence of radiation declines and perturbations begin to enter the MOND regime. Large galaxies could form by $z \approx 10$ (Sanders 2008) and clusters by $z \approx 2$, considerably earlier than in $\Lambda$CDM.  By $z=0$, the voids would become more empty than in $\Lambda$CDM, but otherwise MOND simulations (of collisionless particles, which is of course not the best representation of the baryon fluid) show the same qualitative features of the cosmic web (Llinares et al. 2008). The main difference is in the timing of when structures of a given mass would appear (assembling a large mass early being easier in MOND). This means that MOND is promising in addressing the high-z clusters challenge and Local Void challenge, as well as the bulk flow challenge and high collisional velocity of the bullet cluster (Llinares et al. 2009), due to the much larger than Newtonian MOND force in the structure formation context. What is more, it could allow large massive galaxies to form early ($z \approx 10$) from monolithic dissipationless collapse (Sanders 2008), with well-defined relationships between the mass, radius and velocity dispersion. Consequently, there would be less mergers than in $\Lambda$CDM at intermediate redshifts, but since these mergers would last longer, the number of interacting galaxies could remain unchanged. This could explain the observed abundance of large thin bulgeless disks unaffected by major mergers, and in those rare mergers between spirals, many tidal dwarf galaxies (TDGs) would be formed. This could lead to the possibility that many dwarf galaxies are not primordial but have been formed tidally in these encounters (Kroupa et al. 2010). These populations of satellite galaxies, associated with globular clusters that formed along with them, would naturally appear in closely related planes, thereby providing a natural solution to the satellites phase-space correlation problem (Ibata et al. 2013). To explain why these TDG dwarf satellites appear dark matter-dominated, while TDGs should not if DM is collisionless (see also problem 2.4.d hereabove), MONDian dynamics would be the only possible explanation. What is more, the density-morphology relation for dwarf ellipticals could also find a natural explanation (Dabringhausen \& Kroupa 2012). However, all this would require a covariant parent theory of MOND, which will also have extreme challenges to overcome, and most notably the missing baryon problem that MOND suffers in rich clusters of galaxies and the high third peak in the acoustic power spectrum of the CMB.

\section{MOND-inspired covariant theories: carrying their own problems}

We provide hereafter a non-exhaustive list of the most popular covariant theories that have been proposed to reproduce the MOND phenomenology on galaxy scales. More details and references (as well as a more complete list including proposals based on non-local theories and modified inertia) can be found in Famaey \& McGaugh (2012), as well as in Bruneton \& Esposito-Far\`ese (2007) and Skordis \& Zlosnik (2012). We also point out hereafter some of the main weaknesses of all these approaches. It should be highlighted again that these theories should {\it not} be confused with the MOND paradigm itself, which is not a theory {\it per se} (see e.g. Milgrom 2012).

\subsection{{\bf Scalar-tensor k-essence (RAQUAL)}}

\vspace*{1mm}

{\bf Main ingredients}:  An `Einstein' metric $\tilde{g}_{\mu\nu}$ (with Einstein-Hilbert action), a k-essence scalar field $\phi$, and normal matter from the standard model coupling to a combination of both. \\

\noindent {\bf Action}: The matter couples to the `physical metric':
\begin{equation}
g_{\mu\nu} \equiv e^{2\phi} \tilde{g}_{\mu\nu}
\end{equation}
and the scalar field is given a so-called k-essence action, with no potential and a non-linear, aquadratic, kinetic term:
\begin{equation}
\label{scalaraction}
S_\phi \equiv - \frac{c^4}{2 k^2 l^2 G}\int{d^4 x\sqrt{-\tilde{g}}\, f(X)},
\end{equation}
where $k$ is a dimensionless constant, $l \sim c^2/a_0$ is a length-scale, and $X=kl^2\tilde{g}^{\mu\nu}\phi,_{\mu} \phi,_{\nu}$. The function $f(X)$ yields the MOND dynamics in the approriate limit $X \ll 1$ (Bekenstein \& Milgrom 1984).
\\

\noindent {\bf Main weakness}: Does not enhance gravitational lensing in galaxies because of the conformal relation between the Einstein and physical metrics, a priori enough to reject the theory.

\subsection{{\bf Tensor-Vector-Scalar theory (TeVeS)}}

\vspace*{1mm}

{\bf Main ingredients}:  An Einstein metric $\tilde{g}_{\mu\nu}$ (with Einstein-Hilbert action), a k-essence scalar field $\phi$, a dynamical unit-norm vector field $U_\mu$ allowing a disformal relation between the Einstein and physical metrics (thus allowing to enhance gravitational lensing), and normal matter coupling to a combination of these fields. \\

\noindent {\bf Action}: The matter couples to the physical metric:
\begin{equation}
\label{disformal}
g_{\mu\nu} \equiv e^{-2\phi} \tilde{g}_{\mu\nu} - 2{\rm sinh}(2\phi) U_\mu U_\nu,
\end{equation}
the scalar field is given a k-essence action (Eq.~\ref{scalaraction}), and the vector field (Bekenstein 2004, Skordis 2008):
\begin{equation}
\label{EAaction}
S_U \equiv - \frac{c^4}{16 \pi G}\int{d^4 x \sqrt{-\tilde{g}}\, \left[ K^{\alpha \beta \mu \nu} U_{\beta;\alpha} U_{\nu;\mu} - \lambda (\tilde{g}^{\mu\nu}U_\mu U_\nu + 1) \right] },
\end{equation}
where
\begin{equation}
\label{K}
K^{\alpha \beta \mu \nu} = c_1 \tilde{g}^{\alpha \mu} \tilde{g}^{\beta \nu} + c_2 \tilde{g}^{\alpha \beta} \tilde{g}^{\mu \nu}+ c_3 \tilde{g}^{\alpha \nu} \tilde{g}^{\beta \mu} + c_4 U^\alpha U^\mu \tilde{g}^{\beta \nu}
\label{ci}
\end{equation}
for a set of constants $c_1, c_2, c_3, c_4$, and $\lambda$ is a Lagrange multiplier enforcing the unit-norm.
\\

\noindent {\bf Main weaknesses}: (i) Possible stability issues (especially inside matter), (ii) requires a fine-tuned form of $f(X)$ to evade strong-field observational constraints in the solar system and binary pulsars, (iii) cannot produce a high third peak in the angular power spectrum of the CMB without leading to an exaggerated integrated Sachs-Wolfe (ISW) effect, or without resorting to additional collisionless hot dark matter (appealing to which might overproduce superclusters and massive galaxy clusters as pointed out by Angus \& Diaferio 2011), etc.

\subsection{{\bf Generalized Einstein-Aether theories (GEA)}}

\vspace*{1mm}

{\bf Main ingredients}: Normal matter coupling to the physical metric $g_{\mu\nu}$ (with Einstein-Hilbert action), and a unit-norm dynamical vector field $U_\mu$ also coupling to the physical metric.\\

\noindent {\bf Action}: Zlosnik et al. (2006) showed TeVeS to be expressible as a pure Tensor-Vector theory in the matter frame, with the {\it physical} metric at the same time satisfying the Einstein-Hilbert action and coupling minimally to the matter fields, just like in General Relativity (GR). The GEA approach makes use of this fact to devise a simpler Tensor-Vector action in the matter frame, where the Einstein-Hilbert and matter actions remain precisely as in GR, but with an additional unit-norm vector field (Zlosnik et al. 2007):
\begin{equation}
\label{GEAaction}
S_U \equiv - \frac{c^4}{16 \pi G l^2}\int{d^4 x \sqrt{-g}\, \left[ f(X) - l^2 \lambda (g^{\mu\nu}U_\mu U_\nu + 1) \right] },
\end{equation}
where (see Eq.~\ref{K} and replacing $\tilde{g}^{\mu \nu}$ by $g^{\mu \nu}$)
\begin{equation}
X = l^2 K^{\alpha \beta \mu \nu} U_{\beta,\alpha} U_{\nu,\mu},
\label{XGEA}
\end{equation}
and
\begin{equation}
l =  \frac{(2-K)c^2}{3/2 K^{3/2} a_0}.
\end{equation}
In the static weak-field limit, the unit-norm constraint fixes the vector field in terms of the metric, and from there we have that, in the weak-field limit, $X \propto |\nabla \Phi|^2$.
\\

\noindent {\bf Main weaknesses}: Mostly the same as TeVeS for the CMB and ISW (Zuntz et al. 2010) and for the additional hot dark matter, {\it BUT}, contrary to TeVeS, it can be chosen to approach GR as fast as needed for high accelerations (with no fine tuning) so it avoids all the problems TeVeS has in binary pulsars, and the unacceptably large preferred-frame effects of TeVeS in the solar system.

\subsection{{\bf Bimetric theories (BIMOND)}}

\vspace*{1mm}

{\bf Main ingredients}: Two metrics $g_{\mu\nu}$ and $\hat{g}_{\mu \nu}$, normal matter coupling to $g_{\mu\nu}$, and a new form of `twin matter' coupling to the second metric $\hat{g}_{\mu \nu}$. \\

\noindent {\bf Action}: The heart of this class of theories (Milgrom 2009) is an interaction term between the two metrics, through a function of scalars constructed from $C^{\alpha}_{\mu \nu}/a_0$ where  $C^{\alpha}_{\mu \nu} = \Gamma^{\alpha}_{\mu \nu} -  \hat{\Gamma}^{\alpha}_{\mu \nu}$. For instance, one can write:
\begin{equation}
S \equiv \vphantom{\int}S_{\rm m}[{\rm matter},
g_{\mu\nu}] + \vphantom{\int}S_{\rm m}[{\rm twin \, matter},
\hat{g}_{\mu\nu}] + \frac{c^4}{16\pi G} \int{d^4 x [\alpha \sqrt{-\hat{g}} \hat{R} + \beta \sqrt{-g} R - 2 (g \hat{g})^{1/4} l^{-2} f(X)]},
\label{bimondact}
\end{equation}
where $l = c^2/a_0$, and
\begin{equation}
X = l^2 g^{\mu \nu} (C^{\alpha}_{\mu \beta} C^{\beta}_{\nu \alpha} - C^{\alpha}_{\mu \nu}C^{\beta}_{\beta \alpha}).
\end{equation}
Judiciously chosing the $\alpha$ and $\beta$ parameters yields a whole class of BIMOND theories with various phenomenological behaviors. For instance, in matter-twin matter symmetric versions ($\alpha=\beta=1$), and within a fully symmetric matter-twin matter system, a cosmological constant is given by the zero-point of the function $f$, naturally of the order of 1, thereby naturally leading to $\Lambda \sim a_0^2$ for the large-scale Universe. Matter and twin matter would not interact at all in the high-acceleration regime, and would repel each other in the MOND regime (i.e., when the acceleration difference of the two sectors is small compared to $a_0$), thereby possibly playing a crucial role in the Universe expansion and structure formation.
\\

\noindent {\bf Main weaknesses}: (i) Should still be checked against the existence of ghost modes, (ii) structure formation etc. to be studied in more details, (iii) not clear how these theories can explain the angular power spectrum of the CMB (even when appealing to twin matter).

\subsection{{\bf Dipolar dark matter (DDM)}}

\vspace*{1mm}

{\bf Main ingredients}: Normal matter coupling to the GR physical metric $g_{\mu\nu}$,  dark matter also coupling to the physical metric, but carrying a space-like\footnote{This is to be contrasted with the time-like nature of TeVeS and GEA vector fields in the static weak-field limit} four-vector gravitational dipole moment $\xi^{\mu}$. \\

\noindent {\bf Action}: The dipolar dark matter medium (Blanchet 2007, Blanchet \& Le Tiec 2009) is described as a fluid with mass current $J^\mu = \rho u^\mu$ endowed with the dipole moment vector $\xi^\mu$ (which will affect the total density in addition to the above mass density $\rho$), with the following action:
\begin{equation}
S_{\rm DM} \equiv \int{d^4 x\sqrt{-g}\, [ c^2 (J_\mu \dot{\xi}^\mu -\rho) - W(P) ]},
\label{ddmact}
\end{equation}
where $P$ is the norm of the projection perpendicular to the four-velocity of the polarization field $P^\mu = \rho \xi^\mu$. The Poisson equation in the weak-field limit is then recovered as:
\begin{equation}
-\nabla . ({\bf g} - 4 \pi {\bf P}) = 4 \pi G (\rho_b + \rho).
\label{poissondipolar}
\end{equation}
From there, in order to reproduce the MOND phenomenology in galaxies, one appeals to a `weak-clustering hypothesis', namely the fact that, in galaxies, the dark matter fluid does not cluster much ($\rho \ll \rho_b$) and is essentially at rest (${\bf v} = 0$) because the internal force of the fluid precisely balances the gravitational force, in such a way that the polarization field ${\bf P}$ is precisely aligned with the gravitational one ${\bf g}$, and $g \propto -W'(P)$. At the cosmological level, the monopolar density of the dipolar atoms $\rho$ will play the role of CDM, while the minimum of the potential $W(P)$ naturally adds a cosmological constant term $\Lambda \sim a_0^2$, thus explaining this numerical coincidence and making the theory precisely equivalent to the $\Lambda$CDM model at linear order for the expansion, for large scale structure formation, and for the CMB (naturally explaining the high third peak).
\\

\noindent {\bf Main weaknesses}: (i) Not clear that the weak-clustering hypothesis in galaxies would be the only natural outcome of structure formation within this model, (ii) the weak clustering hypothesis in itself might be problematic for explaining the residual missing mass in galaxy clusters, due to the fact that this residual mass should essentially be concentrated in the central parts of these objects, thus leaving the model to rely on baryonic dark matter or additional hot dark matter to explain galaxy clusters.

\section{Conclusion}

If one is familiar with cosmology and large scale structure, it must seem rather peculiar that anyone would seriously consider alternatives to $\Lambda$CDM based on the MOND phenomenology. But if one is more concerned with the observed phenomenology in a wide range of galaxy data, it seems just as odd to invoke non-baryonic and collisionless CDM together with fine-tuned feedback to explain the appearance of an effective force law that appears to act with only the observed baryons as a source in galaxies (see Sect~2.4 and 3, see also Famaey \& McGaugh 2012 for an extensive review). Accepting this, and building a theory to account for this phenomenology, could perhaps help resolving (see Sect.~3) many of the other challenges (Sects.~2.1--2.3) currently faced by $\Lambda$CDM. Nevertheless, consistent covariant theories that have currently been proposed along these lines are at best effective, and bring with them their own challenges (see Sect.~4). Indeed, the most important aspect before one rejects any model is to have a `simpler' model at hand, that still reproduces the successes of the earlier favored model but also naturally predicts the discrepant data. While MOND, as a paradigm, has {\it a priori} predicted a lot of observations which $\Lambda$CDM cannot explain in galaxies, it is absolutely fair to say that there is currently no alternative which does better {\it overall} than $\Lambda$CDM, and in favor of which Ockham's razor would be. It would however probably be a great mistake to persistently ignore the fine-tuning problems for $\Lambda$CDM and the related uncanny successes of the MOND paradigm on galaxy scales, as they could very plausibly point at a hypothetical better new theory of the dark sector. It would for instance be extremely exciting if one would manage to find a physical connection between the dark energy sector and the galactic MOND phenomenology in the weak-field limit through the $\Lambda \sim a_0^2$ coincidence. In this sense, ideas based on entropic gravity are an interesting line of thought (Verlinde 2011, Ho et al. 2010, Klinkhamer \& Kopp 2011, and others, all somewhat inspired by Milgrom 1999), but one should nevertheless keep in mind the observational challenges that any such new theory of the dark sector could face, especially regarding the third peak of the acoustic power spectrum of the CMB. It is, on the other hand, conceivable that a fully successful alternative theory does not exist, and that the apparent MONDian behavior of galaxies will be explained through small compensatory adjustments of the current $\Lambda$CDM paradigm. But one has to realize that this is far from trivial, and one would have to demonstrate how this could possibly occur. One also has to realize that most currently proposed MOND-inspired effective theories (see Sect.~4) do include new fields, so that the MOND paradigm does not necessarily imply the absence of a dark sector. In any case, the existence of a characteristic acceleration $a_0 \sim \Lambda^{1/2}$ playing various roles in many apparently independent galactic scaling relations is by now an empirically established fact, and it is thus mandatory for \textit{any} successful model of galaxy formation and evolution to explain it, together with all the other challenges mentioned in Sect.~2.

\section*{References}

\begin{thereferences}
\item Angus, G.W., 2009, {\it Mon. Not. R. Astron. Soc.}, {\bf 394}, 527
\item Angus, G.W., Shan, H.Y., Zhao, H.S. and Famaey, B., 2007, {\it Astrophys. J.}, {\bf 654}, L13
\item Angus, G.W. and Diaferio, A., 2011, {\it Mon. Not. R. Astron. Soc.}, {\bf 417}, 941
\item Bekenstein, J., 2004, {\it Phys. Rev. D}, {\bf 70}, 083509
\item Bekenstein, J. and Milgrom, M., 1984, {\it Astrophys. J.}, {\bf 286}, 7
\item Bell, E.F., McIntosh, D.H., Katz, N. and Weinberg, M.D., 2003, {\it Astrophys. J.}, {\bf 585}, L117
\item Blanchet, L., 2007, {\it Class. Quantum Grav.}, {\bf 24}, 3541
\item Blanchet, L. and Le Tiec A., 2009, {\it Phys. Rev. D}, {\bf 80}, 023524
\item Bovill, M.S. and Ricotti, M., 2011, {\it Astrophys. J.}, {\bf 741}, 17
\item Bovill, M.S. and Ricotti, M., 2011, {\it Astrophys. J.}, {\bf 741}, 18
\item Boylan-Kolchin, M., Bullock, J. S. and Kaplinghat, M., 2011, {\it Mon. Not. R. Astron. Soc.}, {\bf 415}, L40
\item Boylan-Kolchin, M., Bullock, J. S. and Kaplinghat, M., 2012, {\it Mon. Not. R. Astron. Soc.}, {\bf 422}, 1203
\item Bruneton, J.-P. and Esposito-Far\`ese, G., 2007, {\it Phys. Rev. D}, {\bf 76}, 124012
\item Clowe, D., et al., 2006, {\it Astrophys. J.}, {\bf 648}, L109
\item Dabringhausen, J. and Kroupa, P., 2012, {\it Mon. Not. R. Astron. Soc.}, arXiv:1211.1382
\item Famaey, B. and McGaugh, S., 2012, {\it Living Reviews in Relativity}, {\bf 15}, 10
\item Frenk, C. and White, S., 2012, {\it Annalen der Physik}, {\bf 524}, 507
\item Gentile, G., Famaey, B., Combes, F., Kroupa, P., Zhao, H.S. and Tiret, O., 2007, {\it Astron. Astrophys.}, {\bf 472}, L25
\item Ho, C. M., Minic, D. and Jack Ng, Y., 2010, {\it Phys. Lett. B}, {\bf 693}, 567
\item Ibata, R., et al., 2013, {\it Nature}, {\bf 493}, 62
\item Kashlinsky, A., Atrio-Barandela, F. and Ebeling, H., 2012, {\it eprint} arXiv:1202.0717
\item Klinkhamer, F. R. and Kopp, M., 2011, {\it Modern Phys. Lett. A}, {\bf 26}, 2783
\item Kroupa, P. et al., 2010, {\it Astron. Astrophys.}, {\bf 523}, A32, (2010)
\item Llinares, C., Knebe, A. and Zhao, H.S., 2008, {\it Mon. Not. R. Astron. Soc.}, {\bf 391}, 1778
\item Llinares, C., Zhao, H.S. and Knebe, A., 2009, {\it Astrophys. J.}, {\bf 695}, L145
\item McGaugh, S.S., Schombert, J.M. and Bothun, G.D., 1995, {\it Astron. J.}, {\bf 109}, 2019
\item McGaugh, S.S. and de Blok, W.J.G, 1998, {\it Astrophys. J.}, {\bf 499}, 66
\item McGaugh, S.S., Schombert, J.M., de Blok, W.J.G. and Zagursky, M.J., 2010, {\it Astrophys. J.}, {\bf 708}, L14
\item Menanteau, F., et al., 2012, {\it Astrophys. J.}, {\bf 748}, 7
\item Milgrom, M., 1983, {\it Astrophys. J.}, {\bf 270}, 365
\item Milgrom, M., 1999, {\it Phys. Lett. A}, {\bf 253}, 273
\item Milgrom, M., 2008, {\it New Astron. Review}, {\bf 51}, 906
\item Milgrom, M., 2009, {\it Phys. Rev. D}, {\bf 80}, 123536
\item Milgrom, M., 2012, arXiv:1212.2568
\item Peebles, P.J.E. and Nusser, A., 2010, {\it Nature}, {\bf 465}, 565
\item Sancisi, R., 2004, {\it IAU Symposium}, {\bf 220}, 233
\item Sanders, R.H., 1998, {\it Mon. Not. R. Astron. Soc.}, {\bf 296}, 1009
\item Sanders, R.H., 2008, {\it Mon. Not. R. Astron. Soc.}, {\bf 386}, 1588
\item Skordis, C., 2008, {\it Phys. Rev. D}, {\bf 77}, 123502
\item Skordis, C. and Zlosnik, T.G., 2012, {\it Phys. Rev. D}, {\bf 85}, 044044
\item Slosar, A., Melchiorri, A. and Silk, J.I., 2005, {\it Phys. Rev. D}, {\bf 72}, 101301
\item Strigari, L.E. and Wechsler, R.H., 2012, {\it Astrophys. J.}, {\bf 749}, 75 
\item Strigari, L.E., 2012, arXiv:1211.7090 
\item Swaters, R., Sancisi, R., van der Hulst, J. and van Albada, T., 2012, {\it Mon. Not. R. Astron. Soc.}, {\bf 425}, 2299
\item Tiret, O. and Combes, F., 2007, {\it Astron. Astrophys.}, {\bf 464}, 517
\item Tiret, O. and Combes, F., 2008, {\it Astron. Astrophys.}, {\bf 483}, 719
\item Tiret, O. and Combes, F., 2008, {\it ASP Conference Series}, {\bf 396}, 259
\item Verlinde, E., 2011, {\it Journal of High Energy Physics}, {\bf 2011}, 29
\item Watkins, R., Feldman, H. A. and Hudson, M. J., 2009, {\it Mon. Not. Roy. Astron. Soc.}, {\bf 392}, 743
\item Zhao, H.S. and Famaey, B., 2012, {\it Phys. Rev. D}, {\bf 86}, 067301
\item Zlosnik, T.G., Ferreira, P.G. and Starkman, G.D., 2006, {\it Phys. Rev. D}, {\bf 74}, 044037
\item Zlosnik, T.G., Ferreira, P.G. and Starkman, G.D., 2007, {\it Phys. Rev. D}, {\bf 75}, 044017
\item Zuntz, J., Zlosnik, T. G., Bourliot, F., Ferreira, P. G. and Starkman, G. D., 2010, {\it Phys. Rev. D}, {\bf 75}, 104015
\end{thereferences}

\end{document}